\documentclass[%
 reprint,
 amsmath,amssymb,
 aps, twocolumn
floatfix,
]{revtex4-1}

\usepackage[%
  colorlinks=true,
  urlcolor=blue,
  linkcolor=blue,
  citecolor=blue
]{hyperref}
\usepackage{graphicx}% Include figure files
\usepackage{dcolumn}% Align table columns on decimal point
\usepackage{bm}% bold math
\usepackage{svg}
\svgsetup{inkscapelatex=false}
\usepackage{scalerel}
\newcommand{\lv}[1]{\textcolor{black}{#1}}

\begin{document}

\preprint{APS/123-QED}

\title{Quantum Inspired Microwave Phase Super-Resolution at Room Temperature}

\author{Leonid Vidro}
\email{leonid.vidro@mail.huji.ac.il}
\author{Liran Shirizly}
\author{Naftali Kirsh}
\author{Nadav Katz}
\author{Hagai S. Eisenberg}

 \affiliation{Racah Institute of Physics, Hebrew University of Jerusalem,
Jerusalem 91904, Israel}

\date{\today}

\begin{abstract}
Quantum metrology has been shown to surpass classical limits of correlation, resolution, and sensitivity. It has been introduced to interferometric Radar schemes, with intriguing preliminary results. Even quantum-inspired detection of classical signals may be advantageous in specific use cases.  Following ideas demonstrated so far only in the optical domain, where practically no thermal background photons exist, we realize room-temperature microwave frequency super-resolved phase measurements with trillions of photons, while saturating the Cramer-Rao bound of sensitivity. We experimentally estimate the interferometric phase using the expectation value of the Parity operator by two methods. We achieve super-resolution up to 1200 times better than the wavelength with 25ns integration time and 56dB SNR.
\end{abstract}

\maketitle

\pagebreak
\section{\label{sec:level1}Introduction}
In Quantum Metrology, non-classical states are used in measurement devices to estimate a physical parameter with enhanced sensitivity. The parameter can be a time interval, an energy gap, or distance, but in most cases, the measurement is achieved by evaluating some phase of an oscillatory system. The most commonly used states are optical states of the electromagnetic field.

Many works have shown the quantum advantages of using squeezed light \cite{KimbleSNL,LigoSensitivity,didier2015heisenberg,bienfait2017magnetic,michael2019squeezing} or photonic NOON states \cite{DowlingNOON,SilberbergNOON,DowlingNOON2}, but as these advantages are very sensitive to losses, their use in real-world applications is severely limited. Thus, an intermediate application regime is defined as \textit{Quantum Inspired} (QIN), where the used states are classical, but the measurement is still quantum mechanical. These applications can have enhanced performance, but cannot exceed classical limits such as sensitivity beyond the Shot Noise Limit (SNL). Nevertheless, they can exhibit lateral super-resolution far beyond the Rayleigh limit, and saturate the classical sensitivity limit of the Cramer-Rao bound over a wide range of parameters.

Previous demonstrations of QIN applications utilized optical photons for high signal-to-noise ratio (SNR) and low background conditions \cite{Cohen:14,Andersen}. Optical photons are advantageous as black body radiation background at room temperatures is negligible in this part of the electromagnetic spectrum. Extension of such ideas to the radio frequency (RF) domain has been suggested in order to improve the sensitivity of Radar systems \cite{DowlingQRadar}. As photon energies in the RF frequency domain are orders of magnitude smaller than in the optical domain, room temperature background can hinder their usefulness.

A central QIN approach, which is also behind the Quantum Radar proposal \cite{DowlingTMSV,AgrawalParity,SRquantumLIDAR}, is to measure the expectation value of the Parity operator instead of the Number operator at the output of an interferometric setup. Both theory and experiments supported the resolution enhancement as well as the optimized sensitivity that such a measurement yields. In previous works, the expectation value of the Parity operator was experimentally estimated for optical fields, where single-photon detectors at room temperature are readily available, and the background temperature is effectively zero. This value was estimated by either a homodyne detection \cite{Andersen} or by a photon-number resolving measurement \cite{Cohen:14}.
In this work, we demonstrate a parity measurement of an interferometric phase between electromagnetic waves in the RF domain, where the thermal background is at least thousands of photons, and the detection of single photons at room temperature is impossible.
\lv{The idea of measuring the Parity operator leads to a convenient detection scheme at the dark port of the interferometer. The parity of the output field is estimated by two different methods, saturating the Cramer-Rao bound of the sensitivity while achieving a super-resolved feature. The width of this feature scales inversely proportional to the square root of the SNR allowing practical advantages in choosing the working point and in resolving phase changes. It also gives the intuitive definition of resolution as the width of the narrowest feature of the measured signal, a more useful sense of resolving changes, when taking noise into account.} In part II we present the theory behind the two methods. The experimental setup and the results are depicted in Parts III and IV, respectively. We conclude with a discussion of these results in Part V.

\section{\label{sec:Theory}Theoretical background}
In what follows, we present a relatively brief theoretical background \lv{that includes the quantum origins of the idea}. For a more detailed background, please see Appendix~\ref{AppTheory}. When making phase sensitive measurements, one chooses to measure and calculate a parameter $M(\phi)$ which is susceptible to the phase. The width of the parameter's function with respect to the phase is the \textbf{resolution} of the measurement
\begin{equation}
    \text{Res}(\phi)=\text{FWHM}(M(\phi)).
\end{equation}
The \textbf{sensitivity} of the phase measurement is derived from the precision of the measurement $\Delta M$ and the slope of the function $M(\phi)$
\begin{equation}
    \Delta\phi=\frac{\Delta M}{\partial M/\partial\phi}.
    \label{eq_def_sen}
\end{equation}
Thus, to have a more sensitive measurement, one would like to measure a low noise parameter and a narrow resolution feature with a steep slope.

When the noise statistics are known, the Cramer-Rao (CR) bound~\cite{CRbound,SensingReview} gives the maximum possible precision of the phase measurement over various measurements and estimators. The classical limit of a phase resolution measurement is half of the used wavelength (such as the Rayleigh limit of imaging resolution), and shot noise imposes the limit for sensitivity. It was shown that the expectation value of the Parity operator $\langle\hat{\Pi}\rangle$ presents super-resolution for phase estimation using coherent states, saturating the shot noise limit when no thermal noise is present~\cite{DowlingTMSV}. $\langle\hat{\Pi}\rangle$ was also shown to exhibit super-sensitivity when using quantum two-mode squeezed vacuum (TMSV) states. The Parity operator is defined as
\begin{equation}
\hat{\Pi} = (-1)^{\hat{N}},
\end{equation}
where $\hat{N}$ is the photon number operator. Its expectation value depends on the probability $P_n$ of measuring $n$ photons, as well as it is proportional to the value at the origin point of the Wigner representation of the quantum state of the electromagnetic field~\footnote{In  units of $\hbar=1$} as function of the field quadratures
\begin{equation}
	\langle\hat{\Pi}\rangle = \sum_{n=0}^{\infty} (-1)^n\cdot P_n = \pi\cdot W(0,0).
\end{equation}

For Gaussian states, the Wigner representation coincides with the classical phase space distribution up to the uncertainty between the two quadratures, which is negligible in this experiment. The classical and quantum Fisher information and the CR bound coincide in this case as well~\cite{QFisherEstimation}.

For a single mode symmetric Gaussian state, $\mu$ is the displacement from the origin (with some predefined reference phase). $N_c=\frac{\mu^2}{2}$ is the average number of photons, while $\sigma$ is its uncertainty.  Note that $\mu=\sqrt{2}|\alpha|$, where $\alpha$ is the conventional displacement of a coherent state and $|\alpha|^2=N_c$.
For a coherent state $\sigma^2 = \frac{1}{2}$ is the minimum possible quantum noise that is symmetric for both quadratures. A coherent state is also a shot noise limited signal. 
For a thermal state $\mu = 0$ and $\sigma^2 = N_{th}+\frac{1}{2}$, where $N_{th}$ is the average number of photons of the thermal state.

For a coherent state superimposed with thermal noise in the same mode, the Wigner representation is just an appropriately scaled convolution of the representations of the two states~\cite{GlauberCoherent,SuperposeWigner}, resulting in
$\frac{\mu^2}{2} = N_c$ and $\sigma^2 = N_{th}+\frac{1}{2}$.
We define a phase space signal-to-noise ratio as
\begin{equation}
    \text{SNR}=\frac{N_c}{N_{th}+1/2}.
\end{equation}
This value coincides with the full definition of signal-to-noise ratio when the thermal noise is much higher than the shot noise. For white noise, such as the noise in our system, the SNR scales down linearly with increased integration time. The expectation value of the Parity operator of such a Gaussian state is
\begin{equation}
    \langle\hat{\Pi}\rangle = \pi\cdot W(0,0) = \frac{1}{2\sigma^2}e^{-\frac{\mu^2}{2\sigma^2}}.
\end{equation}
For an interferometric phase measurement of a coherent state with $N_c$ photons on average and a phase $\phi$, the expectation value of the Parity operator is
\begin{equation}
    \langle\hat{\Pi}\rangle = \frac{1}{2N_{th}+1}e^{-\frac{N_c\cdot \sin^2(\phi/2)}{N_{th}+1/2}}.
    \label{eq_Parity}
\end{equation}The CR bound on measuring an interferometric phase $\phi$ for high SNR is given by:
\begin{equation}
    \Delta\phi_{min} = \sqrt{\frac{2}{\text{SNR}}}= 2\frac{\sigma}{\mu}=\sqrt{2\frac{N_{th}+1/2}{N_c}},
\end{equation} 
which is reduced to the shot noise limit when $N_{th}=0$. 

A direct measurement of the parity is not always available. In fact, sometimes estimating the mean value of the parity from the number of photons is better than from parity values (since the parity value has less information than the number of photons).

By measuring the field quadrature distribution in phase space, $W(0,0)$ can be estimated using a maximum likelihood fit. The error in this estimation can be derived from the Fisher information of the distribution. The sensitivity of the phase measurement approaches the CR bound for large values of SNR.

Another method for estimating $W(0,0)$~\cite{Andersen,Cohen:14} is a projection measurement on $W(0,0)$. For a photon number measurement, this is a projection on the vacuum state, measuring the probability that no photons arrived at all. For a phase space measurement, this would be done by calculating the probability of the measurement being inside some phase space area defined by a threshold radius $a$ from the origin:

\begin{equation}
    \tilde{W}(0,0)\approx \frac{1}{\pi a^2}\int\limits_{0}^{2\pi}d\theta\int\limits_{0}^{a}drrW(r,\theta).
    \label{eq_Parity_Andersen}
\end{equation}

The probability of the measurement being inside this radius is given by the cumulative distribution function (CDF) of the marginal radial distribution in the phase space up to $a$. For a symmetric Gaussian distribution, the marginal radial distribution is Rician.

\begin{figure*}[!t]
\includegraphics [width=0.95\textwidth]{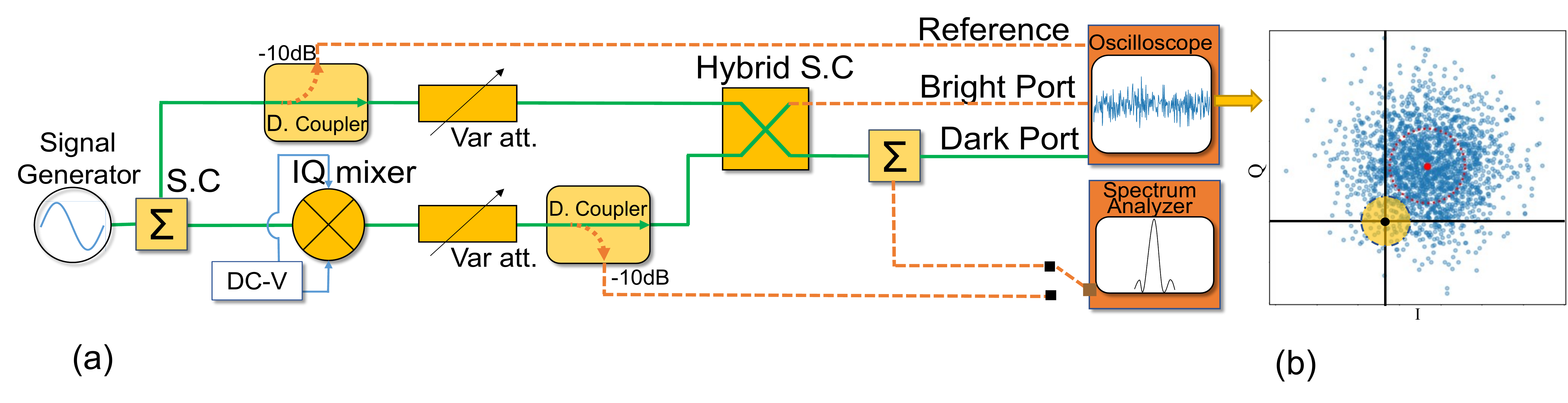}
\caption{\label{Setup} (a) Experimental Mach-Zehnder-like configuration and a sample IQ distribution analysis.
A signal generator is the source of a coherent signal at 4.96GHz. A non-unitary (Wilkinson power divider) splitter-combiner splits the signal into two paths, and a unitary RF Hybrid splitter-combiner interferes the two paths. The two output ports are sampled directly by an oscilloscope. An IQ mixer controls the relative phase at one arm by applying the appropriate DC voltage. A directional coupler at the upper arm provides a phase reference signal. A directional coupler at the lower arm allows the calibration of the working point of the IQ mixer using a spectrum analyzer. (b) Analysis sample: A complex FFT amplitude of each 25ns time series is a point in phase space. The red dot and circle are the mean and standard deviation of the distribution. The Yellow circle centered at the origin designates the area for the threshold analysis. See the main text for more details.}
\end{figure*}

The choice of $a$ affects both the resolution and the sensitivity of the measurement and one can choose a working point that suits the specific requirements of the system. A smaller threshold gives a better resolution, but requires a larger number of measurements. The optimal threshold radius is on the order of $\sigma$. The best possible sensitivity with this method does not saturate the CR bound but asymptotically approaches it up to a factor of $1.26$ calculated numerically, see~\ref{AppTheory:ParityEstimationThreshld}.

% --------------------------
\section{Experimental setup and Analysis}

The experimental setup depicted in Fig.~\ref{Setup} is equivalent to an optical Mach-Zehnder interferometer. A Signal Generator is the source of the RF signal. The first beam-splitter is an RF splitter-combiner, and the latter is an RF Hybrid splitter-combiner~\cite{pozar2011microwave}. At one of the arms, an IQ mixer controls the interferometric phase $\phi$. The output is sampled directly by an oscilloscope, then Fourier analyzed, effectively conducting Homodyne detection. The input signal is also sampled for a phase reference. A Fourier decomposition with phase relative to the reference signal gives the quadratures for a single sample of the phase space distribution. Several sequences of the signal construct the phase space distribution from which $\langle\hat{\Pi}\rangle$ can be estimated.

The signal frequency is 4.96GHz, and the oscilloscope's sampling rate is 20GS/s, well beyond the Nyquist limit. The signal frequency is adjusted to a minimum point of the system spectral noise. The sample duration of 25ns is such that the signal frequency would be exactly on the grid of the discrete Fourier spectrum.

The system was calibrated before measurements for different values of signal power. First, The DC offset of the IQ mixer was found by an automatic minimization procedure using a spectrum analyzer. Then we set port 1 of the Hybrid as the dark port by the same minimization procedure. The latter procedure finds the "zero" interferometric phase and finely adjusts the amplitude to compensate for imbalances in the interferometer, including the extinction ratio of the components, resulting in an overall extinction ratio of 90dB.

Variable digital attenuators are used in each arm in order to control the signal power. The thermal noise of the system is dominated by the oscilloscope noise.
We scanned the phase for five different values of signal power approximately 10dB apart. For each value of the interferometric phase, at each power, the measurement was repeated 200 times to estimate the error.

The SNR of the measurement was controlled by the signal power but can be controlled also by the integration time. See Appendix~\ref{AppIntegration} for more details.

The $W(0,0)$ value is estimated by the two methods described in Sec~\ref{sec:Theory}; by a maximum likelihood fit to the phase space distribution and by using the threshold method, where $\tilde{W}(0,0)$ is evaluated by counting the number of phase space samples within a threshold radius $a$. The phase space distribution was analyzed using 200 samples. The error in estimating $W(0,0)$ scales as the square root of the number of phase space measurements and so is the sensitivity. The presented sensitivity is normalized to a single phase space measurement for better comparison.

\section{Results}

\begin{figure} [t!]

\includegraphics [width=1\linewidth]{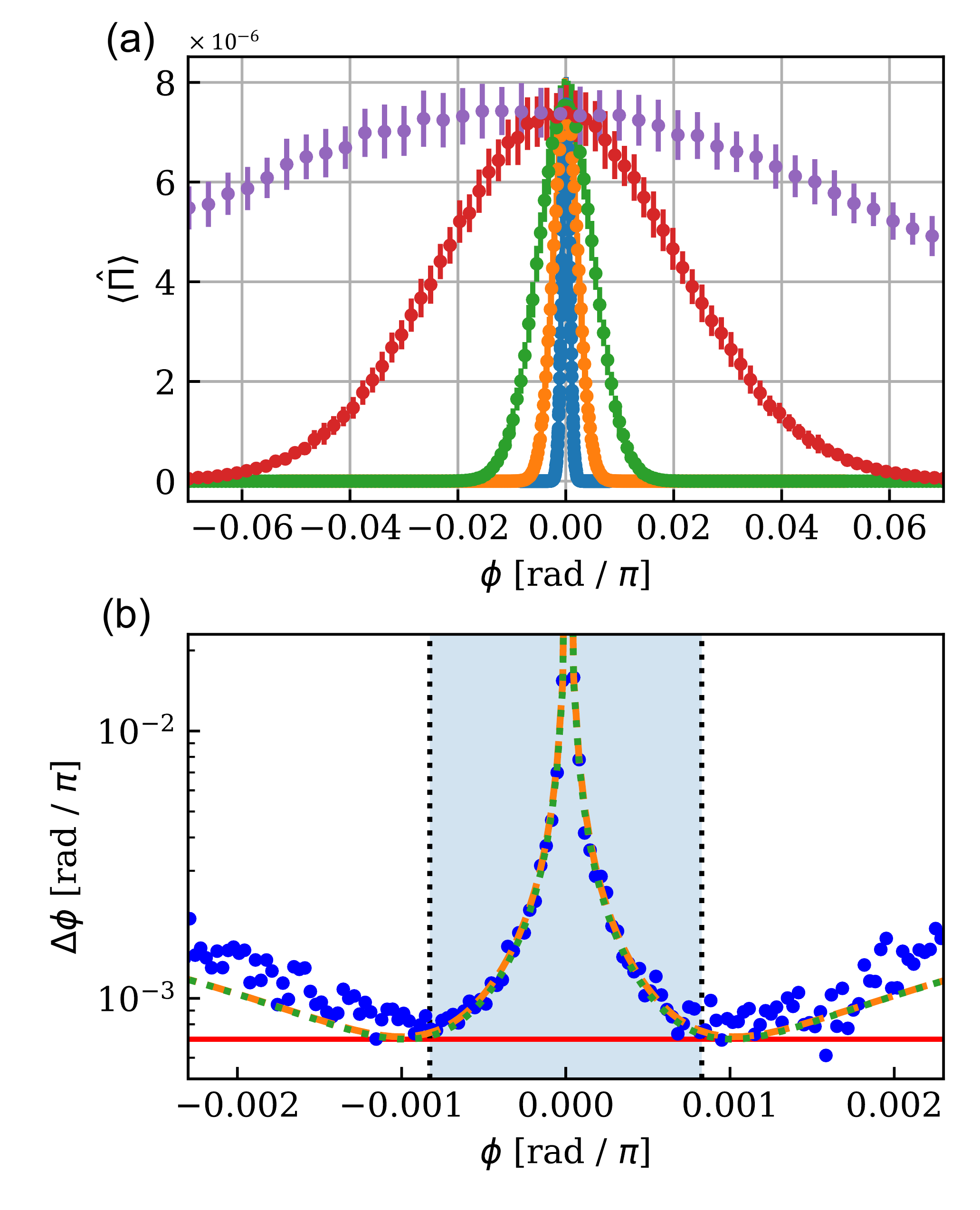}
\caption{Output field parity and its sensitivity, estimated by a maximum likelihood fit. (a) Estimated parity for different SNR values, obtained by the fitting procedure:  Presented curves are for SNR values of 14.8, 26.3, 38.8, 45.9, and 56.1 dB, from wide to narrow trace.
(b) Sensitivity around the minimal output phase, for the SNR=56.1dB measurement, normalized per a single 25ns measurement. Blue dots are values derived from the measured parity, orange and green dashed lines represent the fit and theoretical calculation of the sensitivity, respectively, which are not distinguishable at this scale. The solid red line marks the Cramer-Rao bound. The blue shaded area designated the FWHM range of the parity curve.}
\label{Res_sen_max}
\end{figure}

Figure~\ref{Res_sen_max}(a) presents $\langle\hat{\Pi}\rangle$ as estimated by a maximum likelihood fit over a range of interferometric phases, for several values of signal power. The relevant $N_{c}$ and $N_{th}$ as well as the corresponding SNR were extracted from a fit to Eq.~\ref{eq_Parity}. Errors were estimated by repeating the experiment 200 times for each data point. The maximum value of $\langle\hat{\Pi}\rangle$ depends only on $N_{th}$, thus remaining unchanged over different values of the SNR. $N_{th}$ was also verified independently by a direct measurement, with an agreement to the fit value of $N_{th} = 67.1\pm 0.3\cdot10^{3}$ photons for 25ns integration time. The full-width-at-half-maximum (FWHM) of the parity measurement corresponds to the measurement resolution. The scaling of the FWHM is following $1/\sqrt{\text{SNR}}$ and the highest resolution demonstrated at 56.1dB is 1200 times smaller than the wavelength. The same plot with a logarithmic x-axis is shown in Fig.~\ref{LogxSenResThresh}. Since thermal noise is white, the SNR scales down linearly with longer integration time.

\begin{figure} [t]
{\includegraphics [width=1\linewidth, height=10.66cm ]{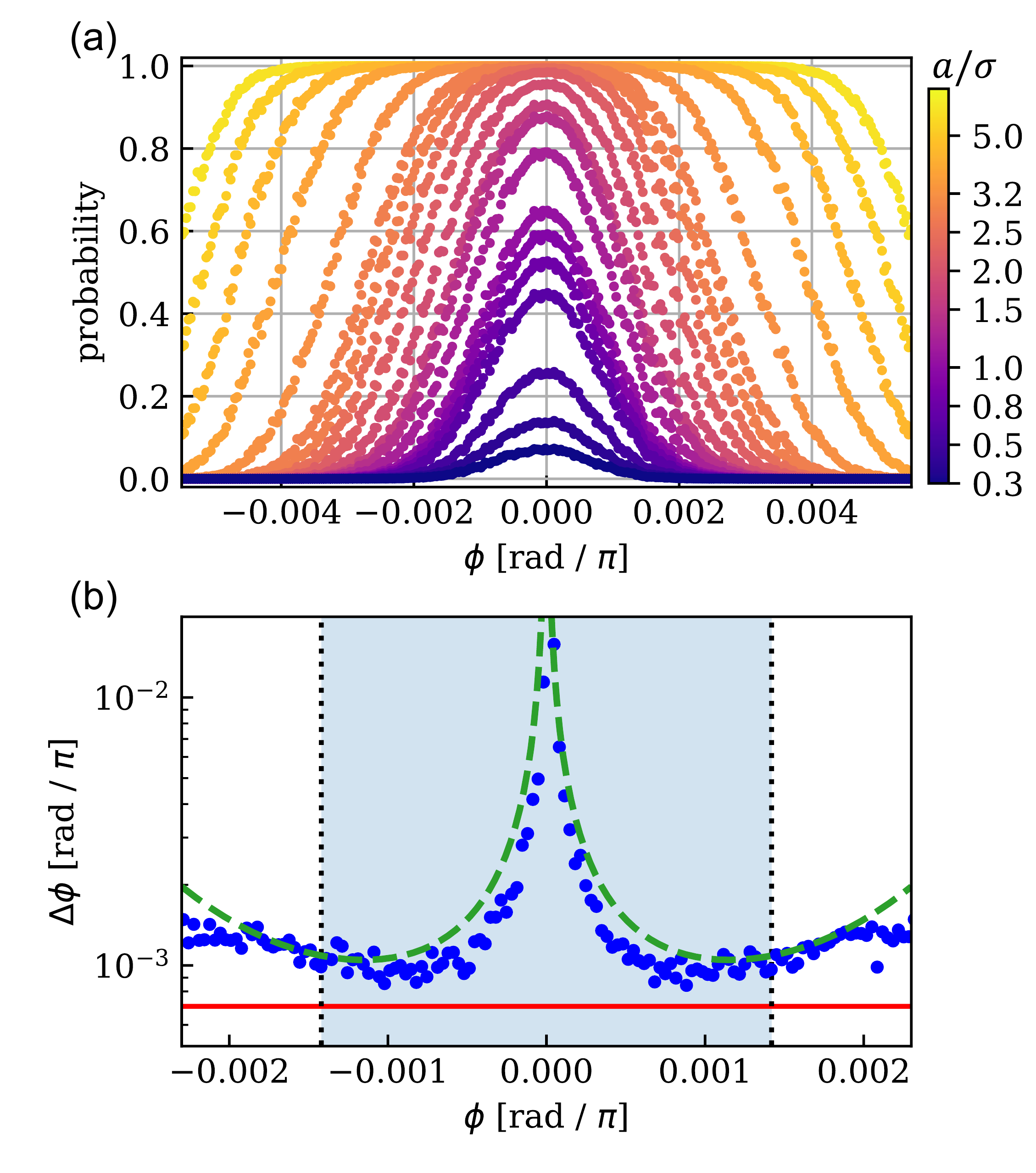}}
\caption{Output field resolution and its sensitivity, estimated by the threshold method. (a) The measured probability of an IQ sample within the threshold area at SNR=56.1dB. The color represents different ratios between the threshold $a$ and the noise $\sigma$ (b) Sensitivity plot derived from the parity measurement for SNR=56.1dB and threshold $a = 1.5\sigma$, normalized per a single 25ns measurement. Blue dots are values derived from the measured parity, green dashed line represents the theoretical calculation of the sensitivity. The solid red line marks the Cramer-Rao bound. The blue shaded area designated the FWHM range of the parity curve.}
\label{Res_sen_thresh}
\end{figure}

Figure~\ref{Res_sen_max}(b) presents the sensitivity as a function of the phase for the highest resolution, normalized per a single phase space measurement (of 25ns). A theoretical curve and a fit to the sensitivity are also shown, both with very good agreement with the measurement central part. The minimum of the fit saturates the CR bound.

Figure~\ref{Res_sen_thresh}(a) presents parity estimation results using the threshold method for the highest value of SNR=56.1dB and different values of threshold radius $a$. When the threshold satisfies $a\gg\sigma$, most of the phase space points are within the threshold radius even when the phase changes. In this case, the estimated parity saturates. When $a\leq\sigma$, the resolution is better. However, many samples are required for appropriate estimation, thus making the measurement less applicable to some cases.

The choice of $a$ affects both the resolution and the minimum possible sensitivity.

Figure~\ref{Res_sen_thresh}(b) presents a sensitivity plot for SNR=56.1dB and $a=1.5\sigma$. The value of $a$ was chosen considering a trade off between sensitivity and resolution.
The sensitivity is normalized per a single phase-space measurement. For more details, see Appendix~\ref{AppTheory:ParityEstimationThreshld}.

\begin{figure} [t]
\centering
\includegraphics [width=1\linewidth]{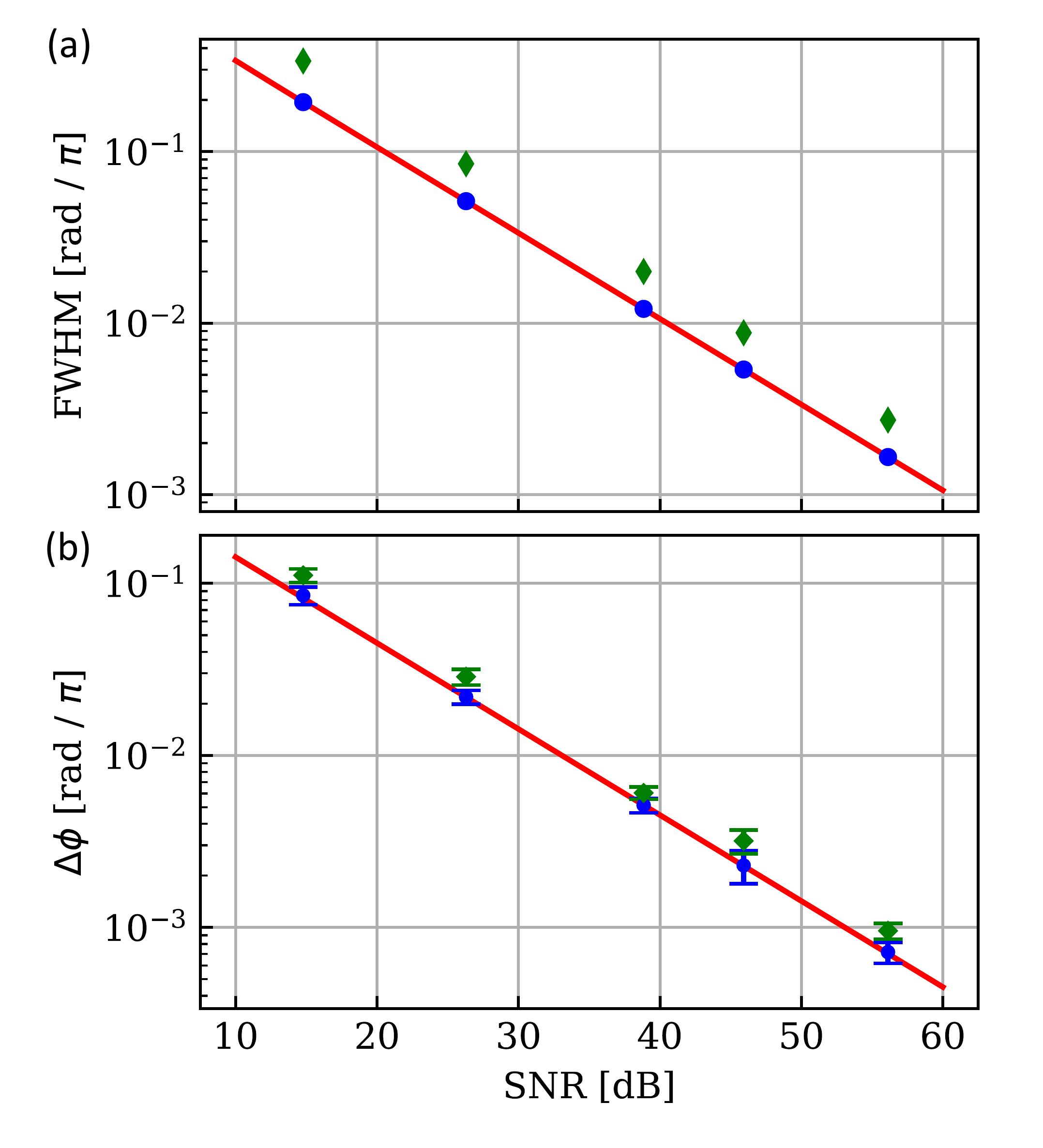}
\caption{(a) Phase resolution of parity measurements. Blue circles are fitted values by the maximum likelihood method for each SNR. Green diamonds are fitted values by the threshold method with $a=1.5\sigma$. The solid red curve is the theory that scales as $1/\sqrt{\text{SNR}}$. Errors are smaller than the symbols. (b) The minimum value of the sensitivity per phase space measurement for the two methods. Blue circles - estimation by maximum likelihood fit. Green diamonds - estimation by the threshold method with threshold $a=1.5\sigma$. Dashed lines represent the corresponding theoretical curves. The solid red line is at the Cramer-Rao bound.}
\label{Res_sen_total}
\end{figure}

Figure~\ref{Res_sen_total} presents the scaling of the resolution and the sensitivity with the SNR, for estimating $\langle\hat{\Pi}\rangle$ by the two methods. The SNR is the figure of merit defining the performance of this measurement at all regimes down to the shot noise limit. In the threshold method, $a/\sigma$ is the second parameter defining the resolution and the sensitivity relative to the CR bound, in addition to the SNR value.

For both methods, the resolution and the sensitivity are of the same order of magnitude, in contrast to the classical method, where the resolution is unaffected by the SNR. Both scale as $1/\sqrt{\text{SNR}}$. The sensitivity of the estimation by the maximum likelihood fit method saturates the CR bound and the threshold method reaches the bound up to a factor of $\sim1.26$ calculated numerically, see~\ref{AppTheory:ParityEstimationThreshld}.

\section{Discussion and Outlook}

We have demonstrated two methods for estimating the expectation value of the Parity operator, both based on the reconstruction of the Wigner representation by an effective Homodyne detection. The Wigner representation reduces to the IQ distribution in this case of Gaussian states and considerable thermal noise. Thus, as the quantum and thermal fluctuations are scaled on a common footing, our work immediately extends to any domain and temperature between the quantum to classical divide.

The first method is based on a maximum likelihood fit of the phase space measurements to a Gaussian distribution. This method saturates the CR bound. It seems that for this method, at least two measurements are required to estimate the maximum likelihood fit, but since the noise in the system should remain constant, it can be calibrated, and a single measurement would suffice to estimate the current mean value of the distribution and to complete the maximum likelihood fit. The threshold method is more straightforward to calculate, though choosing the optimal threshold always requires calibration of the noise. The freedom of choice of the threshold parameter allows choosing a combination of the super-resolution and the sensitivity, where the quality of one comes at the expanse of the other. This trade-off is defined by a single parameter $a/\sigma$. The minimum achievable sensitivity with this method saturates the CR bound only up to a factor of $\sim1.26$. Both methods can have an enhanced resolution by increasing the SNR, e.g., by increasing the integration time. 

\lv{The sensitivity of a measurement describes the certainty in which the phase is estimated. It is influenced by the inherent noise in the physical state that is used for sensing, and by the dependence between the measured quantity and the estimated parameter as in Eq.~\ref{eq_def_sen}.
The resolution property is important when it is desirable to differentiate between two settings of the measured phase. The Rayleigh limit on resolution does not depend on the SNR of the measurement, thus it lacks an important practical aspect. Using the methods presented here, the resolution scales with the SNR the same way as the sensitivity does, giving the resolution its intuitive interpretation in resolving and differentiating between settings. 
}
This work can be extended to free space with phase estimation used for refining both range and direction measurements. Further research includes significantly reducing the microwave system's inherent thermal noise by cooling parts of the setup. Using squeezed states, will result in a super-sensitivity measurement \cite{KimbleSNL}. All of these extensions are also in the framework of Gaussian states.

\section{Acknowledgments}
We would like to thank Dr. Lior Cohen for fruitful discussions and the support of ISF Grants 963.19 and 2323.19.
\appendix 
\counterwithin{figure}{section}

\section{\label{AppTheory} Detailed theoretical background}

For phase-sensitive measurements, the phase is estimated by measuring a parameter $M(\phi)$ which is susceptible to the phase. The width of the parameter's function with respect to the phase is the resolution of the measurement:
\begin{equation}
    \text{Res}(\phi)=\text{FWHM}(M(\phi)).
\end{equation}
In principle, for a noiseless measurement, when the resolution function is unambiguous and reversible over the measurement range - we can identify the phase with infinite precision. When a phase measurement is used for imaging or direction finding, a narrow resolution function enables resolution of smaller features of the image or multiple elements. When a measurement is conducted with noise, the precision of the phase is derived from the precision of the measurement $\Delta M$ and the slope of the function $M(\phi)$:
\begin{equation}
    \Delta\phi=\frac{\Delta M}{\partial M/\partial\phi},
\end{equation} and is called the sensitivity of the measurement. Thus, to have a more senstive measurement, one would like to measure a low noise parameter and a narrow resolution feature with a steep slope.

When the noise statistics are known, the CR bound~\cite{CRbound,SensingReview}, derived from the Fisher information of the signal, gives the maximum possible precision of the phase estimation over various measurements and estimators.
 The classical limit of a phase resolution measurement is half of the used wavelength (such as the Rayleigh limit of imaging resolution), and shot noise imposes the limit for sensitivity. The new possibilities introduced by Quantum Mechanics to phase measurements are achieved by using states of the electromagnetic field with noise lower than the shot noise limit; using entanglement for enhancing correlations and thus signal-to-noise ratio~\cite{Lloyd08,IlluminationDigital,Shapiro}; and also by making measurements of quantum observables (which are also operators in the mathematical representation of quantum mechanics)~\cite{DowlingTMSV}. One such observable is the Parity operator $\hat{\Pi}$, the expectation value of which was shown to present super-resolution for phase estimation using coherent states, saturating the shot noise limit when no thermal noise is present. It was also shown to exhibit super-sensitivity when using quantum two-mode squeezed vacuum (TMSV) states. The Parity operator is defined as:
\begin{equation}
    \hat{\Pi}= (-1)^{\hat{N}},
\end{equation}
where $\hat{N}$ is the photon number operator. It can also be shown that the expectation value of the Parity operator is the value at the origin point $(0,0)$ of the Wigner representation of a quantum state~\cite{WignerTutorial}. The Wigner representation is a quasi-probability distribution of conjugate quantum operators like position and momentum or the quadratures of the electromagnetic field. Conjugate quantum operators follow the Heisenberg uncertainty relation and thus cannot be measured simultaneously without violating the Heisenberg uncertainty principle, which gives the quantum limit of sensitivity~\cite{DowlingTMSV,Hofmann}.

In general, the Wigner representation is very different from an IQ phase diagram, as it can have negative values in restricted areas of the phase space. In addition, a point in phase space cannot be measured directly since this violates the Heisenberg uncertainty principle. However, there are still a variety of quantum phenomena that can be observed and exploited using Gaussian states, which are states for which the Wigner representation is Gaussian and thus can be easily analyzed and measured. Those states include, for instance, coherent states, thermal states, coherent states superimposed with thermal noise, squeezed states and TMSV.
The classical phase space distribution and the Wigner representation coincide in these cases. Moreover, for room-temperature thermal noise, classically measuring the quadratures while ignoring the uncertainty principle has a negligible effect on the results. The classical and quantum Fisher information and CR bound coincide for these states as well~\cite{QFisherEstimation}.
\\
A single mode of the electromagnetic field is similar to a harmonic oscillator with the dimensionless Hamiltonian:
\begin{equation}
    \hat{H} = \frac{1}{2}\hat{X}^2 + \frac{1}{2}\hat{P}^2 = \hat{N}+\frac{1}{2},
\end{equation}
where $X$ and $P$ are the conjugate operators, which are the two quadratures of the electromagnetic field, following the Heisenberg uncertainty principle: $\Delta X\cdot\Delta P \geq \frac{1}{2}$
and also the more subtle $\Delta N\cdot\Delta\phi \geq \frac{1}{2}$.
We present here analysis for a single mode symmetric Gaussian state. The derivation for a general multi-mode Gaussian state is straightforward. The Wigner representation of a symmetric Gaussian state is:
\begin{equation}
W(x,p) = \frac{1}{2\pi\sigma^2}\exp\left(-\frac{(x-\mu \cos(\theta))^2+(p-\mu\sin(\theta))^2}{2\sigma^2}\right),
\end{equation}
where $\mu$ is the displacement of the distribution from the origin with some predefined reference phase $\theta$.
For a coherent state $\sigma^2 = \frac{1}{2}$, is the minimum possible quantum noise that is symmetric for both quadratures. A coherent state is also a shot noise limited signal. $\frac{\mu^2}{2} = N_c$ is the average number of photons. Most commonly, a coherent state is characterized by the complex parameter $\alpha$ for which $|\alpha|^2=N_C=2\mu^2$, its phase is the phase of the coherent signal and the displacement operator is parameterized by it~\cite{GlauberCoherent}. We chose to work here with the definition of $\mu$ since it is more convenient when working with the Wigner representation.
For a thermal state $\mu = 0$ and $\sigma^2 = N_{th}+\frac{1}{2}$, where $N_{th}$ is the average number of photons in the thermal state.
For a coherent state superimposed with thermal noise in the same frequency mode, the Wigner representation is just an appropriately scaled convolution of the Wigner representations of the two states~\cite{GlauberCoherent,SuperposeWigner} resulting in
$\frac{\mu^2}{2} = N_c$ and $\sigma^2 = N_{th}+\frac{1}{2}$. We define a phase space signal-to-noise ratio as:
\begin{equation}
    \text{SNR}=\frac{N_c}{N_{th}+1/2}.
\end{equation}
This value coincides with the full definition of signal-to-noise ratio when the thermal noise is much higher than the shot noise. For white noise such as the noise in our system, the SNR scales down linearly with increased integration time. The expectation value of the Parity operator of such a Gaussian state is:
\begin{equation}
    \langle\hat{\Pi}\rangle = \sum_{n=0}^{\infty} (-1)^n\cdot P_n = \pi\cdot W(0,0) = \frac{1}{2\sigma^2}e^{-\frac{\mu^2}{2\sigma^2}}.
\end{equation}
$P_n$ is the probability of measuring $n$ photons.
For an interferometric phase measurement of a coherent state with $\mu_0^2/2$ total number of photons and an interferometric phase $\phi$, at the output of one interferometer arm $\mu = \mu_0\cdot \sin(\phi/2)$, and the number of photons is:
 \begin{equation}
     \mu^2/2 = N_c\cdot \sin^2(\phi/2) = \mu_0^2/2\cdot\sin^2(\phi/2)
 \end{equation}
and the phase
\begin{equation}
    \theta = \phi/2.
\end{equation}
The expectation value of the Parity operator thus follows:
\begin{equation}
    \langle\hat{\Pi}\rangle = \pi\cdot W(0,0) = \frac{1}{2N_{th}+1}e^{-\frac{N_c\cdot \sin^2(\phi/2)}{N_{th}+1/2}}.
    \label{App_eq_Parity}
\end{equation}

From the phase space distribution, for cases in which the SNR is large, the standard deviation of the phase can be easily calculated to be $\Delta\theta=\frac{\sigma}{\mu}$ by changing to polar coordinates, thus the CR bound on measuring an interferometric phase $\phi$ for high SNR is given by:
\begin{equation}
    \Delta\phi_{min} = 2\frac{\sigma}{\mu_0}=\sqrt{\frac{2}{\text{SNR}}}=\sqrt{2\frac{N_{th}+1/2}{N_c}},
\end{equation}
which is reduced to the shot noise limit when $N_{th}=0$.

\subsection{\label{AppTheory:DirectParityMeasurement} Direct measurement of $\hat{\Pi}$}
 Since the Parity operator is $\hat{\Pi}=(-1)^{\hat{N}}$, a direct measurement of the Parity operator is possible if a photon number measurement is available. This is achievable more easily at room temperature environment in the optical regime, but also in the RF domain with cryogenics. In this case, its mean value could be calculated simply by ensemble averaging. The error of such a measurement is the standard deviation of the Parity operator and the calculated sensitivity of the measurement is:
\begin{equation}
    \Delta\phi = \frac{\sqrt{\frac{1}{2N_{th}+1}e^{\frac{4N_c\sin^2(\phi/2)}{2N_{th}+1}}-1}}{|\frac{N_c}{2N_{th}+1}\sin(\phi)|}.
\end{equation}

\subsection{\label{Theory:ParityEstimationML} Estimating $\langle\hat{\Pi}\rangle$ by Maximum likelihood}
A direct measurement of parity is not always available. In fact, sometimes estimating the mean value of the parity from the number of photons is better than from parity values (since the parity value has less information than the number of photons).
By measuring the field quadrature distribution in phase space, the $W(0,0)$ value can be estimated using a maximum likelihood fit.
The Fisher information for estimating the mean and variance of a 2-D symmetric Gaussian distribution by a maximum likelihood fit is given by:
 \begin{equation}
     \mathcal{I}(\mu,\sigma^2) =
     \begin{pmatrix}
     \frac{1}{\sigma^2} & 0\\
     0 & \frac{1}{\sigma^4}
     \end{pmatrix}.
 \end{equation}
Using this, the error of estimating the value at the origin, normalized to a single phase space measurement is:
 \begin{equation}
     \pi\cdot\Delta W(0,0) = \Delta \langle\hat{\Pi}\rangle = \frac{1}{2\sigma^2}e^{-\frac{\mu^2}{2\sigma^2}}\cdot \sqrt{1+\frac{\mu^2}{\sigma^2}}.
 \end{equation}
The sensitivity of this estimation is given by:
 \begin{equation}
     \Delta\phi = 
     \sqrt{(\cot^2(\phi/2)+\left(\frac{N_{th}+1/2}{N_c}\right)^2\frac{2}{\sin(\phi)}},
 \end{equation}
the minimum value of this sensitivity is given by:
 \begin{equation*}
 \begin{aligned}
     \Delta\phi_{min} 
     & = \sqrt{\frac{2}{\text{SNR}^2}+2\sqrt{\frac{1}{\text{SNR}^2}\left(\frac{1}{\text{SNR}^2}+1\right)}}
\\
&     \xrightarrow[SNR\to\infty]{} \sqrt{\frac{2}{\text{SNR}}},
 \end{aligned}
 \end{equation*}
which reduces to the CR bound for large values of SNR.

\begin{figure}[t]
  \centering
  \includegraphics [width=\linewidth]{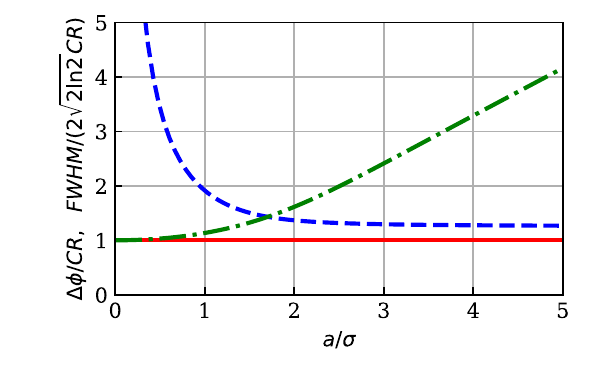}
  \caption{Resolution and sensitivity as a function of the threshold $a$. Green dot-dashed line represents resolution, and Blue dashed line is for the sensitivity. Solid red line represents the Cramer-Rao (CR) bound. The x-axis is normalized per standard deviation of the noise $\sigma$, and the y-axis is per one CR bound.}
  \label{SenResThresh}
\end{figure}

\subsection{\label{AppTheory:ParityEstimationThreshld} Estimating $\langle\hat{\Pi}\rangle$ by Threshold analysis}
Another method for estimating $W(0,0)$~\cite{Cohen:14,Andersen} would be an approximation to a projection measurement on $W(0,0)$. For a photon number measurement, this would be a projection on the vacuum state, measuring the probability that no photons arrived at all. For a phase space measurement, this would be done by calculating the probability of the measurement being inside some phase space area defined by a threshold radius $a$ from the origin:
\begin{equation}
    \tilde{W}(0,0)\approx\frac{1}{\pi a^2}\int\limits_{0}^{2\pi}d\theta\int\limits_{0}^{a}drrW(r,\theta).
    \label{Appeq_Parity_Andersen}
\end{equation}
The probability of the measurement being inside this radius is given by the cumulative distribution function (CDF) of the marginal radial distribution in the phase space up to $a$. For a symmetric Gaussian distribution, the marginal radial distribution is Rician.
The smaller the threshold the better the estimation in the limit of many measurements, but a larger number of measurements would be required. The choice of the threshold radius $a$ also defines different performance attributes for resolution and sensitivity. ~Fig.~\ref{SenResThresh} Shows the trade-off between resolution and sensitivity for different values of $a$.
This plot is universal when $a$ is normalized to $\sigma$ (defined by the noise) and when the sensitivity is normalized to the CR bound (defined by the SNR). The resolution here is normalized to $2\sqrt{2ln2}$ times the CR bound. This factor relates the resolution defined by FWHM in contrast to the sensitivity defined by standard deviation. With this normalization, the trade-off between resolution and sensitivity could be better appreciated.
The minimum possible resolution with this method matches that of the maximum likelihood method.
However, The minimum possible sensitivity does not saturate the CR bound but asymptotically reaches it up to a factor of $\sim1.26$ which was calculated numerically. The balance point of similar performance of both resolution and sensitivity is at $a/\sigma\approx1.7$.

This threshold method is a projection measurement or a Bernoulli trial with two outcomes. One of them is the probability of having a sample inside the radius $a$, which resembles the estimate of $\langle\hat{\Pi}\rangle$ given by Eq.~\ref{Appeq_Parity_Andersen}

\begin{equation}    p=\int\limits_{0}^{2\pi}d\theta\int\limits_{0}^{a}drrW(r,\theta).
    \label{Appeq_Probability_Andersen}
\end{equation}

\begin{figure}[t]
  \center{
  \includegraphics [width=\linewidth]{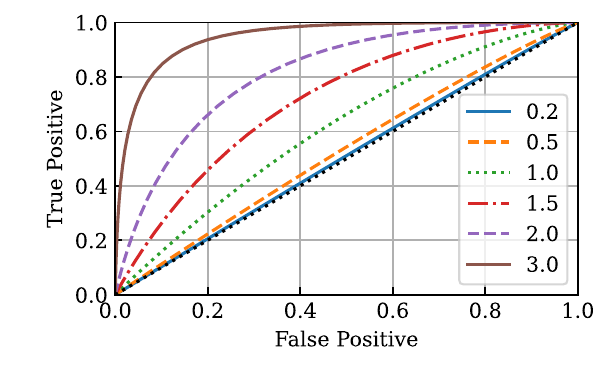}}
  \caption{ROC curve for several Acceptable Deviations (AD) in units of the Cramer-Rao bound (CR). From straight to curved, the curves correspond to 0.2AD/CR (blue) up to 3AD/CR (brown). Each curve is parametrized by the detection threshold $a/\sigma$. In these normalized units the figure is universal.}
  \label{ROC}
\end{figure}

Using this equation we can calculate the resolution for a given phase space distribution. The error in calculating the Bernoulli probability $p$ is just $\Delta p=\sqrt{p}$ per phase space sample point (and is reduced as the square of the number of sample points) and so the sensitivity can be calculated.
A different combination of resolution and sensitivity for different applications might be useful. For a given $N_c$ and $N_{th}$ which define the distribution, it is possible to choose the best working point in terms of combined resolution and sensitivity, as shown in Fig.~\ref{SenResThresh}. Another performance analysis can be made for applications in which phase locking is required. For such applications, one needs to define the precision within which one wants to be locked. Suppose one wants to be locked on the phase to within an acceptable deviation (AD) $\Delta \phi_l$ from perfect locking. This gives us two binary thresholds. $\Delta\phi_l$ is the locked\textbackslash unlocked binary threshold for the true state of the system and $a$ in a binary threshold of our estimation. Now we can calculate the probabilities of actual conditions versus predicted conditions. For example, in Radar applications, it is common to plot a Receiver Operating Characteristic (ROC) curve, which is the true positive alerts versus false positive alerts probabilities. In our case, a true positive occurrence is when $\phi<\Delta \phi_l$ and we received a sample within the threshold $a$. A false positive occurrence is when $\phi>\Delta \phi_l$ but we still received a sample within the radius $a$. Fig.~\ref{ROC} shows parametric ROC curves for various values of locking precision $\Delta \phi_l$ parametrized with the detection threshold $a$. The locking precision is stated in terms of the CR bound. When $a$ is stated in terms of $\sigma$ this is a universal plot. Fig.~\ref{ROC} is another tool to choose a working point for the threshold method.

\begin{figure}[t]
\centering
  \includegraphics [width=1\linewidth]{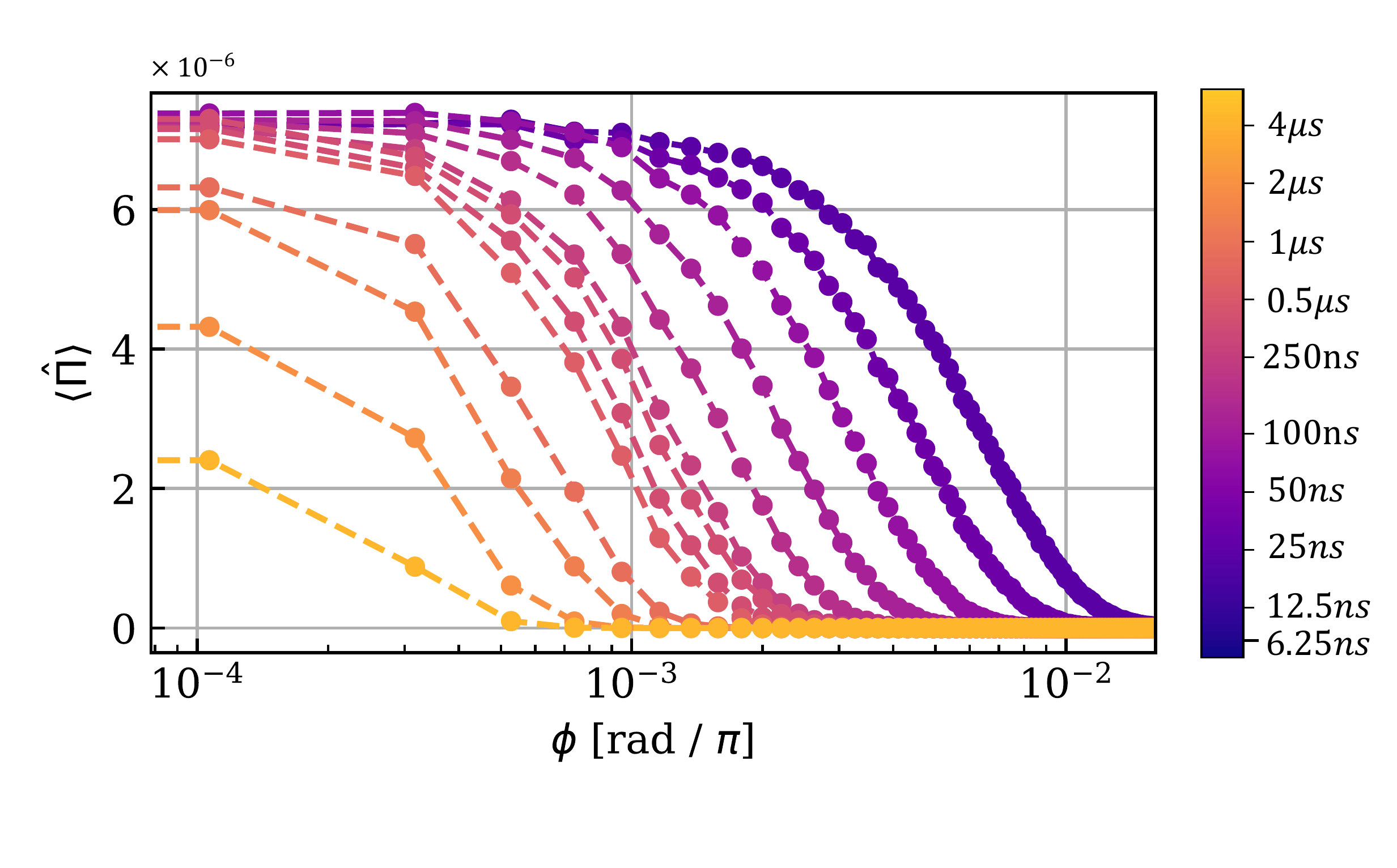}
\caption{Resolution feature of $\langle\hat{\Pi}\rangle$ using the maximum likelihood method for different integration times at SNR = 45.9dB. As the integration time grows, first the width is just reduced as the SNR increases. At a certain point, the noise becomes comparable to the remaining signal power due to the imperfect extinction ratio, and the height of the feature is reduced as well. 
}
\label{dif_len}
\end{figure}

\begin{figure}[t]
  \centering
  \includegraphics [width=\linewidth]{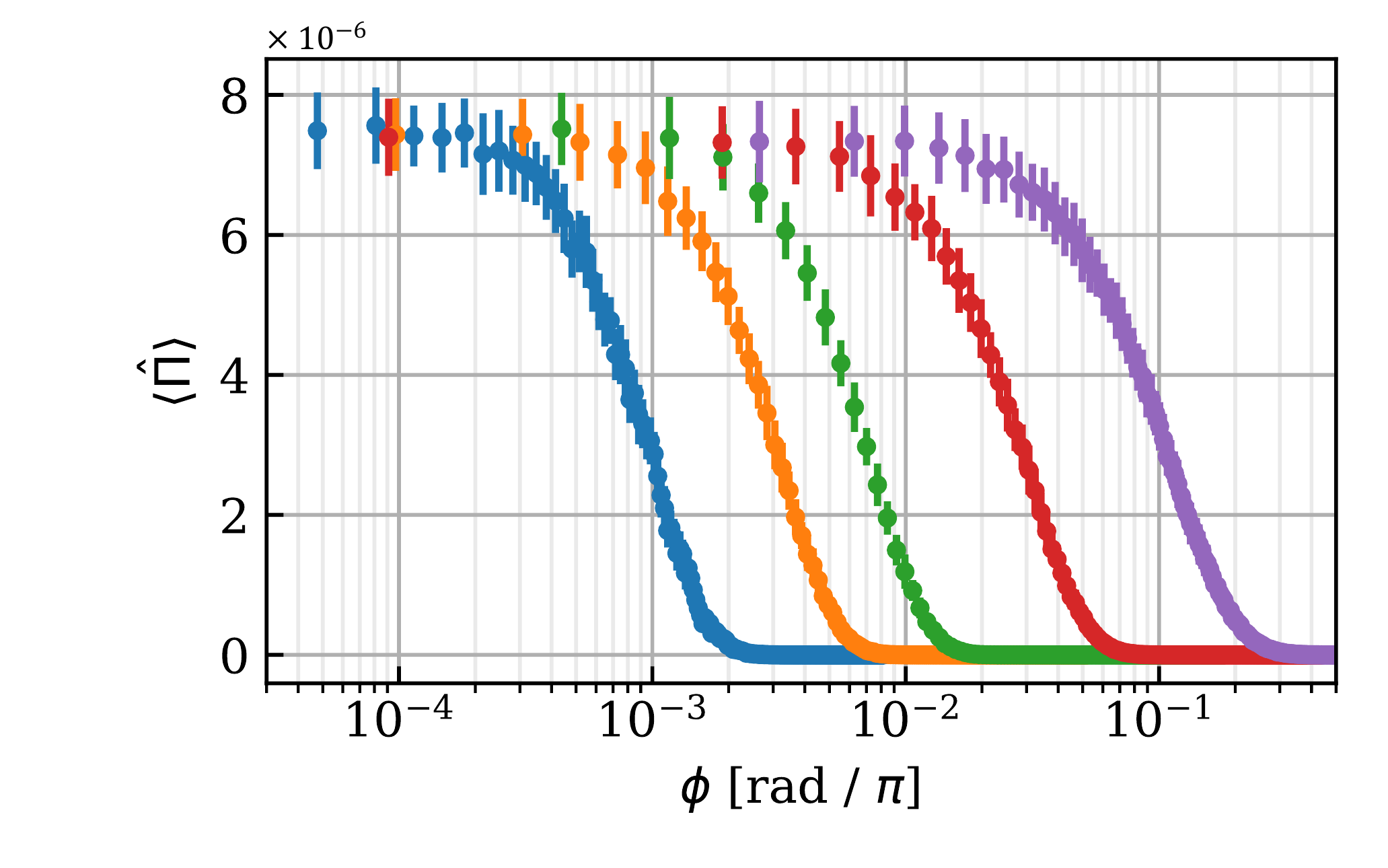}
  \caption{Parity calculated using the maximum likelihood method with logarithmic x-axis, the width scales linearly with the SNR.}
  \label{LogxSenResThresh}
\end{figure}

\section{\label{AppIntegration} Controlling the SNR by integration time}
The FWHM of $\langle\hat{\Pi}\rangle$ follows $1/\sqrt{\text{SNR}}$ scaling. The different SNR values in the main paper were achieved by controlling the signal power with a constant integration time of 25ns. Higher SNR and higher resolution can also be obtained by a longer integration time with constant signal power.
When considering the energy within a certain integration time, $N_c$ increases linearly with integration time, however $N_{th}$ does not change. This happens due to the fact that the frequency mode of the photons is defined up to the inverse of integration time, thus when the integration time grows, the overall average noise energy increases linearly but the bandwidth of the mode decreases linearly leaving $N_{th}$, the overall average number of photons in the mode, constant (since the thermal noise is white). We can also analyze the system in units of power, treating $N_c$ and $N_{th}$  as the average number of photons per second. In this case, $N_c$ does not change with the integration time but rather $N_{th}$ decreases linearly since it is normalized per a certain time interval. The quadrature noise in this case scales as $\sqrt{\text{Hz}}$. The overall effect, however, does not change since the SNR changes in the same way in both pictures.

Increasing the integration time is limited in practice by the setup's extinction ratio, which was 90dB in our setup. Such a high extinction ratio was achieved by controlling the losses in both arms of the interferometer and minimizing the dark port output power. The resolution feature of $\langle\hat{\Pi}\rangle$ for several integration times with constant power is shown in Fig.~\ref {dif_len}.
Ideally, increasing the integration time results only in narrowing the resolution, and the desired SNR can be achieved with longer integration times. However, when the noise of the distribution becomes comparable to the leakage power in the interferometer due to the imperfect extinction ratio, it lowers the maximum value of the resolution feature and reduces the quality of the measurement.

Figure~\ref{LogxSenResThresh} shows $\langle\hat{\Pi}\rangle$ as estimated by a maximum likelihood fit for several values of signal power at a constant integration time of 25ns. It is plotted with logarithmic x-axis, which shows clearly the dependence of the resolution width on SNR.

% \nocite{*}
\bibliography{apssamp}

\end{document}